\documentclass[12pt]{article}%
\usepackage{amsmath}
\usepackage{amsfonts}
\usepackage{amssymb}
\usepackage{graphicx}
\usepackage{hyperref}%
\ifx\pdfoutput\relax\let\pdfoutput=\undefined\fi
\newcount\msipdfoutput
\ifx\pdfoutput\undefined\else
\ifcase\pdfoutput\else
\ifx\paperwidth\undefined\else
\ifdim\paperheight=0pt\relax\else\pdfpageheight\paperheight\fi
\ifdim\paperwidth=0pt\relax\else\pdfpagewidth\paperwidth\fi
\fi\fi\fi
\begin{document}

\title{Dark matter from primordial metric fields and the term (grad
g\_\{00\})\symbol{94}2}
\author{M. Chaves\\\textit{Universidad de Costa Rica}\\\textit{San Jos\'{e}, Costa Rica}\\mchaves@fisica.ucr.ac.cr}
\date{June 29, 2011}
\maketitle

\begin{abstract}
It is a well-known truism, inspired on the general theory of relativity, that
gravity gravitates. Here we suggest the possibility that dark matter may be
caused by the gravitation of the metric. At first sight this seems impossible
since the gravitational fields in galaxies and cumuli are so weak that it
would seem that second order terms are negligible. Nevertheless, the general
theory of relativity tells us that the gravitation due the metric is given by
(grad g\_\{00\})\symbol{94}2. Thus, a metric field g\_\{00\} varying fast in
the space directions could make a sizeable contribution to the gravitational
field despite being a weak field. As a plausible source of such a field
consider that during reheating the inflaton field disintegrates into
radiation. Those quantum decays that involve higher energies and momenta will
produce pockets of metric fields with rapid change in time and space. The
expansion of the universe and dissipative processes (including the emission of
gravitational waves) eventually result in basically stationary pockets of
classical g\_\{00\} field varying rapidly in space that have collapsed along
with matter into structures like galaxies and cumuli. These pockets should
gravitate precisely by the term mentioned above. They are classical fields and
are reminiscent of the cosmic fields that are scattered in our universe.

\end{abstract}

\textit{The dark matter hypothesis. }The dark matter hypothesis has been very
useful to explain the discrepancy between mass as estimated from its
luminosity and as inferred from its gravitational effect. Of course, it is
also possible that this discrepancy can be due to the use of incorrect
dynamical theories, which in this case would be the general theory of
relativity. Many alternative models of dynamics have been presented in the
last few years to explain the discrepancy, some covariant, some
not.$^{\cite{CCCT,MOND}}$ However, several observational results of the last
few years, based on the techniques of gravitational lensing, have greatly
substantiated the dark matter hypothesis. Dark matter has been, basically,
directly observed,$^{\cite{proof DM}}$ and even dark matter cosmological
structures.$^{\cite{scaffolding}}$ There are several hypothesis as to what is
dark matter is, but its actual composition is still a mystery. In this paper
we argue for yet another new possible candidate for dark matter.

\textit{Gravity gravitates. }People often say that the nonlinearity of the
equations of general relativity imply that gravity gravitates. Here we raise
the question: could it be, perhaps, that the gravitation of gravity in
galaxies and clusters constitutes dark matter? We begin with a brief study of
the pertinent equations of the general theory of relativity.

Within general relativity the geodesic equation%
\[
\frac{d^{2}x^{\lambda}}{d\tau^{2}}+\Gamma_{\mu\nu}^{\lambda}\frac{dx^{\mu}%
}{d\tau}\frac{dx^{\nu}}{d\tau}=0.
\]
serves as an equation of motion for the force of gravity. We first calculate
the equation of motion in the Newtonian limit. We consider the particularly
simple case of bodies with small velocities interacting solely through a
rather weak gravitational field. We can then take the metric to be diagonal
and neglect its time dependence. The geodesic equation becomes%
\begin{equation}
\frac{d^{2}x^{i}}{dt^{2}}=-\Gamma_{00}^{i}=\frac{1}{2}g^{ii}g_{00,i}%
\approx\frac{1}{2}g_{00,i},\text{ }i=1,2,3. \label{motion}%
\end{equation}
Taking the gauge%
\begin{equation}
g_{00}=-1-2\psi, \label{metric}%
\end{equation}
where $\psi$ is interpreted as the gravitational potential, the equation
(\ref{motion}) can be written in the form $\mathbf{\ddot{x}}=-\mathbf{\nabla
}\psi,$ Newton's equation for a particle in a gravitational field. (The
signature of the metric is $(-+++).)$

In taking the Newtonian limit of general relativity, one approximates matter
in the galaxy by a fluid with a density $\rho_{M}(t,\mathbf{x})$ and
negligible pressure. Then the stress tensor has only one nonzero component,
$T_{00}=\rho_{M}.$ Einstein's field equation can be written in the alternative
form $R_{\mu\nu}=8\pi G(T_{\mu\nu}-\frac{1}{2}g_{\mu\nu}T),$ from which we can
immediately infer that $R_{00}=R^{\rho}{}_{0\rho0}=4\pi G\rho_{M},$ or%
\[
\Gamma^{\rho}{}_{00,\rho}-\Gamma^{\rho}{}_{0\rho,0}+\Gamma^{\rho}{}_{\rho\eta
}\Gamma^{\eta}{}_{00}-\Gamma^{\rho}{}_{0\eta}\Gamma^{\eta}{}_{\rho0}=4\pi
G\rho_{M}.
\]
Let us take the Newtonian limit of this equation, except that we keep the two
terms on the left the side of this equation that involve the $g_{00}$:%
\begin{equation}
-\frac{1}{2}\nabla^{2}g_{00}-\frac{1}{4}(\nabla g_{00})^{2}=4\pi G\rho_{M}.
\label{2-order}%
\end{equation}
Usually the second term on the left is omitted because it is second order in
$g_{00}$ and thus seems negligible for the gravitational field of a galaxy or
cumulus. If we substitute in this equation using (\ref{metric}) we can write
it in the form%
\begin{equation}
\nabla^{2}\psi=4\pi G\rho_{M}+(\nabla\psi)^{2}, \label{Poisson}%
\end{equation}
which has the interpretation that both matter density $\rho_{M}$ and
$(\nabla\psi)^{2}$ are sources of gravity, that is, both matter and gravity gravitate.

Notice the term $(\nabla\psi)^{2}$ always gravitates with a positive sign,
just like matter. Notice, too, that its contribution goes as the square of the
\textit{gradient} of the potential. This is a very important point, because
even if the gravitational field $\psi$ is small, its gradient does not have to
be so. If in the galaxy's halo exist pockets of rapidly varying gravitational
field $\psi(x,y,z),$ they could contribute significantly to the average
galactic gravitational field $\psi(r),$ where $r$ is the radial coordinate of
the galaxy. Thus the second-order term in (\ref{2-order}) does not have to be
smaller that the first-order term, and it could even be larger.

These pockets of rapidly varying gravitational field are not gravitational
waves, but large volumes of classical, almost stationary but rapidly varying
in space, metric fields. In the same way that there exist cosmic magnetic
fields scattered throughout the universe, there can exist these metric fields.
Both are primordial in origin, a point we shall develop with respect to the
gravity pockets.

These pockets differ from other dark matter candidates in that they are not
elementary particles and so cannot be observed by experiments involving
particle detectors. Like matter, they gravitate and possess inertia. They do
not interact with matter at all except gravitationally. In particular, they
are impervious to the pressure exerted by cosmic thermal radiation.

\textit{Plausibility of the existence of gravity pockets. }A plausible
primordial origin for gravity pockets is during the reheating of the universe
that is supposed to occur after the end of inflation. During reheating, the
state of the inflaton $\varphi$ field falls down a steep potential curve
$V(\varphi)$ and eventually oscillates at the concave bottom of the potential.
All through this fall and the oscillation at the bottom of the potential the
inflaton is decaying into radiation and producing a lot of entropy, and
consuming in the process a potential energy density just a few orders of
magnitude smaller than $E_{P}^{4}$. The inflaton has a large coherent
component $\varphi$ whose size is model-dependent, but is roughly of the order
of Planck's energy: $\varphi\sim$ $E_{p}=G^{-1/2}$. However, according to
quantum field theory, there should exist a perturbative incoherent component,
call it $\varphi^{\prime}$, which decays into radiation. These decays are
stochastic and produce radiation with varying energy and momentum. The ones
that have large energy and momentum can produce powerful local modifications
in the metric. The usual first-order cosmological perturbation
theory$^{\cite{MFB}}$ does not work in describing these events and the
calculation has to be carried out to second-order, since both orders can be of
comparable sizes for the reasons stated above.

The stress tensor for the inflaton is%
\[
T_{\mu\nu}=\varphi_{,\mu}\varphi_{,\nu}-\frac{1}{2}g_{\mu\nu}g^{\alpha\beta
}\varphi_{,\alpha}\varphi_{,\beta}-g_{\mu\nu}V(\varphi),
\]
and Einstein's equation during inflation and reheating is%
\begin{equation}
R_{\mu\nu}-\frac{1}{2}g_{\mu\nu}R=8\pi G\left(  T_{\mu\nu}+...\right)  =8\pi
E_{P}^{-2}\left(  T_{\mu\nu}+...\right)  , \label{Einstein}%
\end{equation}
where the ellipsis represents other stress tensors that may be relevant. The
value of $T_{00}$ is usually taken to be a few orders of magnitude smaller
than $E_{P}^{4}$ to avoid entering a quantum gravity regime (where we cannot
calculate),$^{\cite{W}}$ so the right-side of (\ref{Einstein}) is a few orders
of magnitude smaller than $E_{P}^{2}.$ We conclude that the differentiated
metrics on the left side of (\ref{Einstein}) are equated to very strong
changes on the right during reheating, when the potential $V(\varphi)$ is
rapidly diminishing. The volumes surrounding the more energetic
disintegrations of the perturbative inflaton become pockets of rapidly
changing metric. During this process radiation and gravitational waves are
emitted. Eventually the expansion of the universe and dissipative processes
conspire to reduce the time dependence of the metric and result in basically
stationary gravity pockets of classical $g_{00}$ metric field.

\textit{Calculation of virial velocity curves assuming gravity pockets in a
weak galactic field. }We want to prove that this explanation for dark matter
is consistent with the observed virial velocity curves, and also to show how
gravity pockets can be treated in a way similar to matter. Regarding the
virial velocity curves $v(r),$ it is well-known that for a star outside the
visual edge of a galaxy (or a galaxy outside the visual edge of a cluster)
they often take a flat horizontal shape.

Consider a galaxy with an average weak, radial field. The $g_{00}$ field
within a pocket is also weak, but it varies rapidly in all three directions
$x,y,$ and $z$ and gravitates as $\left(  \nabla\psi\right)  ^{2}$ so its
contribution does not have to be negligible. Assume the gravity pockets are
numbered by $i$ and have volumes $V_{i}$. Then the mass-like contribution of a
pocket is given by%
\begin{equation}
P_{i}=\int_{V_{i}}(\nabla\psi)^{2}dV_{i}.\label{pocket}%
\end{equation}
(The value of the integral does not depend on where we locate the origin of
the coordinates we use to perform it.) It is possible to conclude from
(\ref{Poisson}) and (\ref{pocket}), by means of Gauss' Theorem, that if we
have a gravity pocket $P$ and a matter lump of mass $M,$ both situated at the
origin, their combined gravitational potential is%
\begin{equation}
\psi_{M,P}(r)=-\frac{GM}{r}-\frac{P}{4\pi r}.\label{potential}%
\end{equation}
We see that gravity pockets gravitate in a manner very similar to matter lumps.

To proceed with the calculation of the virial velocity curves we need to know
the dependence of matter density $\rho_{M}(r)$ and the gravity pocket density
$\rho_{P}(r)$ on the galaxy's radius $r.$ The present picture of galaxy
formation is that, after reheating, radiation and dark matter permeate the
universe. Dark matter is attracted by the density fluctuations of primordial
quantum origin, and begins collapsing and facilitating the formation of
structure. At about the onset of matter domination matter begins to take part
in the growth of cosmic structures. If we assume that, in the formation of a
galaxy, gravity pockets and matter collapse as a dissipationless gas and
undergo violent relaxation,$^{\cite{LB}}$ then the resulting matter and
gravity pocket densities should decrease as $\rho_{M}\propto\rho_{P}\propto
r^{-2}.$ (This result does not hold for the region near the galactic center).
If we assume that $\rho_{P}\propto r^{-2},$ then the virial velocity curve
will be flat, as we shall see.

In the case of elliptical galaxies the same fall for the density $\rho_{P}$
can be obtained from the use of the Hubble-Reynolds law%
\[
I=\frac{I_{0}r_{H}^{2}}{(r+r_{H})^{2}},
\]
for surface brightness $I,$ where $r_{H}$ is usually taken to be quite smaller
than the galactic radius $r$. For radii not near the center of the galaxy we
conclude that $I\propto r^{-2}.$ If the surface brightness $I$ is proportional
to the density brightness $\rho_{B}$, and the density brightness is
proportional to the matter density $\rho_{M}$, and the matter density is
proportional to the pocket density $\rho_{P}$, then again $\rho_{P}\propto
r^{-2}.$

At any rate let this be our choice for how $\rho_{P}$ falls with $r$ and with
it let us go back to the problem of the virial curves. Assume a galaxy with a
matter lump density $\rho_{M}$ and a gravity pocket density $\rho_{P},$ both
densities possessing, in the average, spherical symmetry. From the spherical
symmetry and (\ref{potential}) it is possible to conclude for this galaxy,
that%
\begin{equation}
\nabla^{2}\psi=4\pi G\rho_{M}+\rho_{P}.\label{Newtonian}%
\end{equation}
Now let $\mathbf{F=-\nabla}\psi$ be the gravitational force. Then, from
(\ref{Newtonian}) and Gauss' Theorem, we obtain:%
\[
F=-\frac{G}{r^{2}}\int_{r}\rho_{M}dV-\frac{1}{4\pi r^{2}}\int_{r}\rho_{P}dV,
\]
where we have assumed a spherical volume of integration with a radius $r.$ For
a radius $r>$ $r_{E},$ where $r_{E}$ is the radius of the galaxy's edge, the
gravitational force would be given by:%
\begin{equation}
F=-\frac{GM}{r^{2}}-\frac{P}{4\pi r^{2}}-\frac{1}{r^{2}}\int_{r_{E}}^{r}%
\rho_{P}(r^{\prime})r^{\prime2}dr^{\prime},\label{force}%
\end{equation}
where the gravity pockets integral has been split in two parts, one
integrating up to the edge, and another from the edge on. We are wondering if
the third term on the right in (\ref{force}) does not fall as rapidly as
$r^{-2}.$ We take then $\rho_{P}=Cr^{-2}$ and and use this function in the
third term to see how fast it falls with $r$:%
\begin{equation}
-\frac{1}{r^{2}}\int\rho_{P}r^{2}dr=-\frac{C}{r^{2}}\int dr^{\prime}=-\frac
{C}{r}.
\end{equation}
Since this third terms falls more slowly (it can also be argued that the curve
has to be continuous), let us simply take $F=-C/r.$ The virial equation for a
particle in a central force is $2\left\langle T\right\rangle =-m\left\langle
Fr\right\rangle ,$ or%
\[
\left\langle mv^{2}\right\rangle =-\left\langle mFr\right\rangle =\left\langle
m\frac{C}{r}r\right\rangle =mC,
\]
from which we conclude that $v^{2}=C,$ that is, the velocity curve outside the
edge is flat.

\textit{Final remarks.} We posit a new candidate for dark matter, our old
friend the $g_{00}$ metric component. We have discussed how it is plausible
that, during reheating, quantum decays (the ones involving higher energies and
momenta) of the perturbational inflaton $\varphi^{\prime}$ result in pockets
of rapidly varying metric. The expansion of the universe and dissipative
processes eventually modify the pockets of $g_{00}$ field so that they are
basically stationary but rapidly varying in the space directions. They
gravitate according to the positive-definite term $(\nabla g_{00})^{2}$ (as
required by the general theory of relativity) so that even for weak
gravitational fields the value of $(\nabla g_{00})^{2}$ does not have to be
necessarily negligible if $g_{00}$ is varying fast enough. The result, large
pockets of rapidly-varying classical $g_{00}$ field, are analogous to the
cosmic magnetic fields that exist throughout our universe.

\textbf{Acknowledgement.} The author would like to thank R. Oreamuno for
productive conversations during the writing of this paper.

\end{document}